\documentstyle[11pt,newpasp,psfig,twoside]{article}
\index{instructions}
\index{guidelines}

\def\edcomment#1{\iffalse\marginpar{\raggedright\sl#1\/}\else\relax\fi}
\reversemarginpar

\begin{document}
\title{New Pulsars from Arecibo Drift-Scan Searches}
\author{M. A.~McLaughlin,
 D. R.~Lorimer,
 D. J.~Champion}
\affil{University of Manchester, Jodrell Bank Observatory, Macclesfield, UK}
\author{K. Xilouris}
\affil{University of Virginia, Charlottesville, VA, USA}
\author{Z. Arzoumanian}
\affil{USRA/LHEA, NASA-GSFC, Greenbelt, MA, USA}
\author{D. C. Backer}
\affil{University of California, Berkeley, CA, USA}
\author{J. M. Cordes}
\affil{Cornell University, Ithaca, NY, USA}
\author{A. S. Fruchter}
\affil{Space Telescope Science Institute, Baltimore, MD, USA}
\author{A. N. Lommen}
\affil{Franklin \& Marshall College, Lancaster, PA, USA}
\begin{abstract}
We report on new pulsars discovered in Arecibo drift-scan data.
Processing of 2200 deg$^{2}$
of data has resulted in the detection of 41 known and 12 new pulsars.
New pulsars include two millisecond pulsars, one solitary and one binary recycled pulsar,
 and one pulsar with  very unusual
pulse profile morphology and complex drifting subpulse behavior.
\end{abstract}

In McLaughlin et al. (2003) we describe the data collection and analysis procedures
for this survey.
At the completion of this analysis,
we have redetected 41 known pulsars and 12 new pulsars. These new pulsars
are J0152+09 ($P$ = 915~ms), J0546+24 ($P$ = 2.8~s), J0609+21 ($P$ = 56~ms), J0815+09 ($P$ = 845~ms),
J1453+19 ($P$ = 5.8~ms), J1504+21 ($P$ = 3.3~s), J1746+22 ($P$ = 3.5~s), J1823+06 ($P$ = 753~ms),
J1829+24 ($P$ = 41~ms), J1944+09 ($P$ = 5.2~ms), J2007+09 ($P$ = 458~ms) \& J2045+09 ($P$ = 395~ms).

\smallskip
\noindent
{\bf PSR~J0609+2130:} From a year of timing, we establish that this pulsar is recycled and is a
solitary neutron star (Lorimer et al. 2003).
 It and 
 PSR~J2235+1506 (Camilo et al. 1993) both lie in a part of the $P-\dot{P}$ diagram 
consisting almost completely of binary pulsars. We
hypothesize that these unique objects are the end-products of massive
binary systems that disrupted following the mass-transfer phase due to
the supernova explosion of the companion star. We 
 should be able to measure a proper motion within 1--2~years,
offering us insights into the formation of these isolated pulsars.
  
\smallskip
\noindent
{\bf PSR~J0815+09:} As shown in Figure~1,
the profile morphology is not easily explained by
standard models. In addition, drifting is seen in all components with
$P_3 = 15 P_1$ and $P_2 = 0.05 P_1$; the sense of drift differs in different components,
with no sense reversals.
These properties are consistent at multiple
epochs and frequencies. No timing irregularities are
apparent; the derived age and magnetic field are
10~Myr and
$8.1\times10^{11}$~G, respectively.

\begin{figure}
\psfig{file=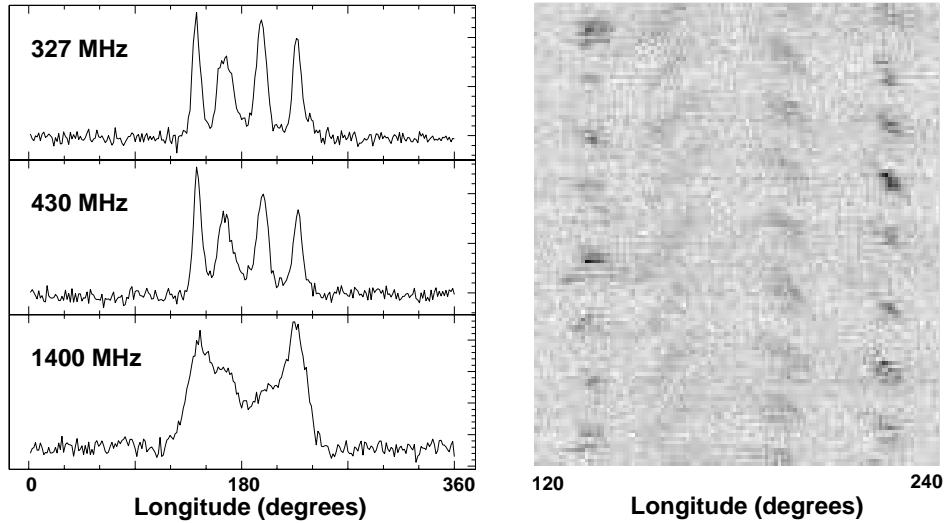,width=5.2in}
\caption{Left: Multi-frequency pulse profiles of J0815+09.
Right: Sequence of individual pulses at 327~MHz.}
\end{figure}

\smallskip
\noindent
{\bf PSR~J1829+24:}
Timing observations of this pulsar over a 4-month timespan reveal a binary orbit of
1.2 days, with eccentricity of 0.14 and projected semi-major axis of 7.2~light~seconds.
 The minimum companion mass is 1.28~$M_{\sun}$,
making the companion most likely a neutron star. We derive a coalesence time of
$\sim 50$~Gyr. This pulsar is quite similar to 
PSR~J1518+4904 (Nice et al. 1996), but has a more compact orbit and hence an expected $\dot{\omega}
\sim 30$ times larger.

\end{document}